\documentclass[aps,prb,twocolumn,epsfig,reprint,amsmath,groupedaddress,amssymb,float]{revtex4-1}

\usepackage{graphics}
\usepackage{graphicx}
\usepackage{epsfig}
\usepackage{subfigure}
\usepackage{amssymb}
\usepackage[dvips]{color}

\begin{document}

\preprint{}

\title{Structure and composition tunable superconductivity, band topology and elastic response of
hard binary niobium nitrides Nb$_2$N, Nb$_4$N$_3$ and Nb$_4$N$_5$}

\author{K. Ramesh Babu$^1$ and Guang-Yu Guo$^{1,2}$ }

\affiliation{$^1$Department of Physics, National Taiwan University, Taipei 10617, Taiwan\\
$^2$Physics Division, National Center for Theoretical Sciences, Taipei 10617, Taiwan\\
}

\date{\today}

\begin{abstract}
We perform a systematic \textit{ab initio} density functional study of
the superconductivity, electronic and phononic band structures, electron-phonon coupling and
elastic constants of all four possible structures of niobium nitride $\beta$-Nb$_2$N as well as
Nb-rich $\gamma$-Nb$_4$N$_3$ and N-rich $\beta^\prime$-Nb$_4$N$_5$. First of all,
we find that all four structures of $\beta$-Nb$_2$N are superconductors
with superconducting transition temperatures ($T_c$) ranging from 0.6 K to 6.1 K,
depending on the structure. This explains why previous experiments reported contradicting $T_c$
values for $\beta$-Nb$_2$N. Furthermore, both $\gamma$-Nb$_4$N$_3$ and $\beta^\prime$-Nb$_4$N$_5$
are predicted to be superconductors with rather high $T_c$ of 8.5 K and 15.3 K, respectively.
Second, the calculated elastic constants and phonon dispersion relations show
that all the considered niobium nitride structures
are mechanically and dynamically stable. Moreover, the calculated elastic moduli demonstrate
that all the niobium nitrides are hard materials with bulk moduli and hardness being comparable to
or larger than the well-known hard sapphire.
Third, the calculated band structures reveal that the nitrides possess both type I and type II
Dirac nodal points and are thus topological metals.
Finally, the calculated electron-phonon coupling strength, superconductivity and
mechanical property of the niobium nitrides are discussed in terms of their
underlying electronic structures and also Debye temperatures.
The present \textit{ab initio} study thus indicates that $\beta$-Nb$_2$N,
$\gamma$-Nb$_4$N$_3$ and $\beta^\prime$-Nb$_4$N$_5$ are hard superconductors with nontrivial band
topology and are promising materials for exploring exotic
phenomena due to the interplay of hardness, superconductivity and nontrivial band topology.
\end{abstract}

\maketitle

\section{INTRODUCTION}
Transition metal nitrides (TMNs) are well known for their refractory characteristics such as high mechanical strength,
hardness, high melting point, excellent thermal stability and resistance to corrosion and oxidation.
These superior properties make them as promising materials for many practical applications,
such as wear-resistance surfaces, high pressure and magnetic storage devices, and cutting tools \cite{LEToth, Zerr}.
Furthermore, TMNs are good metallic conductors and some of them exhibit
superconductivity \cite{Ralls, Gavaler, Skokan, Keskar, Hemley, Wu}. Interestingly,
these materials were also found to possess nontrivial band topology \cite{Alex, Bian, KR}.

Among all the TMNs, the binary niobium nitride systems are of particular interest because they exist
in a variety of crystal structures with outstanding electronic and superconducting properties \cite{Simak, Olifan, Zou, Ma}.
At ambient pressure, the following crystalline structures of the niobium nitrides (see Table I)
are known to exist: (i) cubic $\alpha$-NbN,
(ii) hexagonal $\beta$-Nb$_2$N, (iii) tetragonal $\gamma$-Nb$_4$N$_3$, (iv) cubic $\delta$-NbN,
(v) hexagonal $\varepsilon$-NbN, (vi) hexagonal WC-NbN, (vii) tetragonal $\beta^\prime$-Nb$_4$N$_5$,
(viii) hexagonal $\delta^\prime$-NbN and (ix) hexagonal $\varepsilon^\prime$-Nb$_5$N$_6$.

One interesting feature of these nitride systems is that the Nb atoms are connected with N atoms through strong covalent bonds,
thus resulting in superior mechanical properties compared to the metal carbides and oxides \cite{Kumada}.
The superconductivity of these niobium nitrides depends on both the Nb/N ratio and
the crystal structure \cite{Oya} (see, e.g., Table I). For example, $\delta$-NbN, $\beta$-Nb$_2$N, $\gamma$-Nb$_4$N$_3$
and $\beta^\prime$-Nb$_4$N$_5$ are known to be superconductors while hexagonal $\delta^\prime$-NbN
and $\varepsilon^\prime$-Nb$_5$N$_6$ structures do not exhibit superconductivity down to 1.8 K \cite{KR,Oya}.
Because of their relatively high superconducting transition temperatures and high hardness,
the $\delta$ and $\gamma$ phases of NbN have found applications
in superconducting radio frequency circuits \cite{Sugahara,Radparvan}, Josephson junction qubits \cite{Inomata,Senapat},
terahertz wave detection hot-electron-bolometer \cite{Kang}, superconducting nanowire single-photon
detectors \cite{Lee} and also in the fabrication of superconducting quantum interference devices
(SQUIDS) \cite{Noya, Ren, Tang}. In addition, nitrogen rich structures $\beta^\prime$-Nb$_4$N$_5$
and $\varepsilon^\prime$-Nb$_5$N$_6$ are candidates for supercapacitor applications \cite{Feng}.

However, the superconductivity as well as mechanical and electronic properties of many niobium nitrides
have been rather poorly understood.
In particular, a wide range of superconducting transition temperatures ($T_c$) have been
reported for the $\beta$-phase Nb$_2$N ($\beta$-Nb$_2$N)~\cite{Gavaler,Skokan,Gajar,Jena,Kalal}.
For example, Gavaler {\it et al.} reported that $\beta$-Nb$_2$N
has a $T_c$ value between 8.6 K - 12.1 K \cite{Gavaler}. Skokan {\it et al.}\cite{Skokan} reported
that the thin films of mixed phases of cubic-NbN and hexagonal $\beta$-Nb$_2$N exhibit
two step resistance drop at 9 K and at 2 K. Gajar {\it et al.} \cite{Gajar} reported  the transformation
of Nb into hexagonal $\beta$-Nb$_2$N which is superconducting only below 1.0 K.
Very recently, Kalal {\it et al.}\cite{Kalal} claimed that the hexagonal $\beta$-Nb$_2$N (P6$_3$/mmc)
films have the electron-phonon interaction dominanted superconductivity  with a $T_c$ of 4.74 K.
Clearly, all these experimental studies on superconductivity of $\beta$-Nb$_2$N contradict with each other.

On the other hand, we note that at least four crystalline structures (see Table I and Fig. 1) have been reported
for $\beta$-Nb$_2$N \cite{Tera,Savage,Norlund,Zha,Daul,Chihi,Riyan}.
Guard \textit{et al.}\cite{Savage} reported that $\beta$-Nb$_2$N adopts a W$_2$C type structure
with space group P$\bar{3}$m1. However, Christensen \cite{Norlund} reported that $\beta$-Nb$_2$N
has a $\varepsilon$-Fe$_2$N type structure with space group P$\bar{3}$1m. Besides, $\beta$-Nb$_2$N
also exists in P6$_3$mmc space group \cite{Tera}. Recent \textit{ab initio} random structure search
also predicted that $\beta$-Nb$_2$N can exist in an orthorhombic structure with Pnnm space group \cite{Zha}.
Unfortunately, unlike niobium nitrides with other Nb/N ratios such as NbN where
one structure is labelled as one phase (Table I), all the structures of Nb$_2$N
have been labelled as the $\beta$-Nb$_2$N. It is well-known that the superconductivity
and physical properties of a solid are determined by its crystal structure, as we have recently
demonstrated for NbN~\cite{KR}. Consequently, we believe that the contradicting superconducting
properties reported for $\beta$-Nb$_2$N are caused
by the fact that it has several different structures, as for NbN (Table I).
In this work, therefore, we perform a systematic {\it ab initio} study of
the superconducting and also other physical properties of $\beta$-Nb$_2$N in
all the four possible structures. Furthermore, to study how the superconductivity
depends on the Nb/N ratio, we also consider Nb-rich $\gamma$-Nb$_4$N$_3$ and N-rich $\beta^\prime$-Nb$_4$N$_5$.
Both $\gamma$-Nb$_4$N$_3$ and $\beta^\prime$-Nb$_4$N$_5$ crystallize in
the tetragonal NaCl-type $\delta$-NbN structure, respectively, by removal of half
of either nitrogen or niobium atoms in alternating planes along the $c$-axis \cite{GBrauer}.
They are also superconductors with quite high $T_c$ values (7.8 $\sim$ 16.0 K)~\cite{Oya,Kaiser,Merch}.

Materials that exhibit both superconductivity and nontrivial band topology are excellent candidates
to study the fascinating phenomena such as topological superconductivity and Majorana Fermions \cite{sczhang}.
In recent years, there is indeed a growing interest in the search for materials
where superconductivty coexists with nontrivial band topology \cite{Konig, Sharma}. In the binary Nb-N systems,
the electronic structure \cite{Matt, DJChadi, KSchwarz, Litinskii}, mechanical \cite{Wen, Olifan}, phonon and
superconducting properties \cite{BPalanivel, Isaev} of niobium mononitride (Nb/N = 1) have recently been extensively studied,
and as a result, Dirac and Weyl nodal points have been predicted in several structures
of NbN such as cubic $\delta$-NbN, hexagonal $\varepsilon$-NbN, $\delta^\prime$-NbN, and WC-NbN
by the \textit{ab initio} calculations~\cite{Alex, Bian, KR}.
However, for either Nb-rich or N-rich niobium nitrides (i.e., niobium nitrides with Nb/N ratios different from 1),
no theoretical studies on the band topology and superconductivity have been reported.
Finally, the mechanical properties of either Nb-rich or N-rich niobium nitrides have been much
less investigated and consequently remain poorly understood. This would certainly
hinder their technological applications as hard superconductors.

\begin{figure*}
\centering
\includegraphics[width = 160mm]{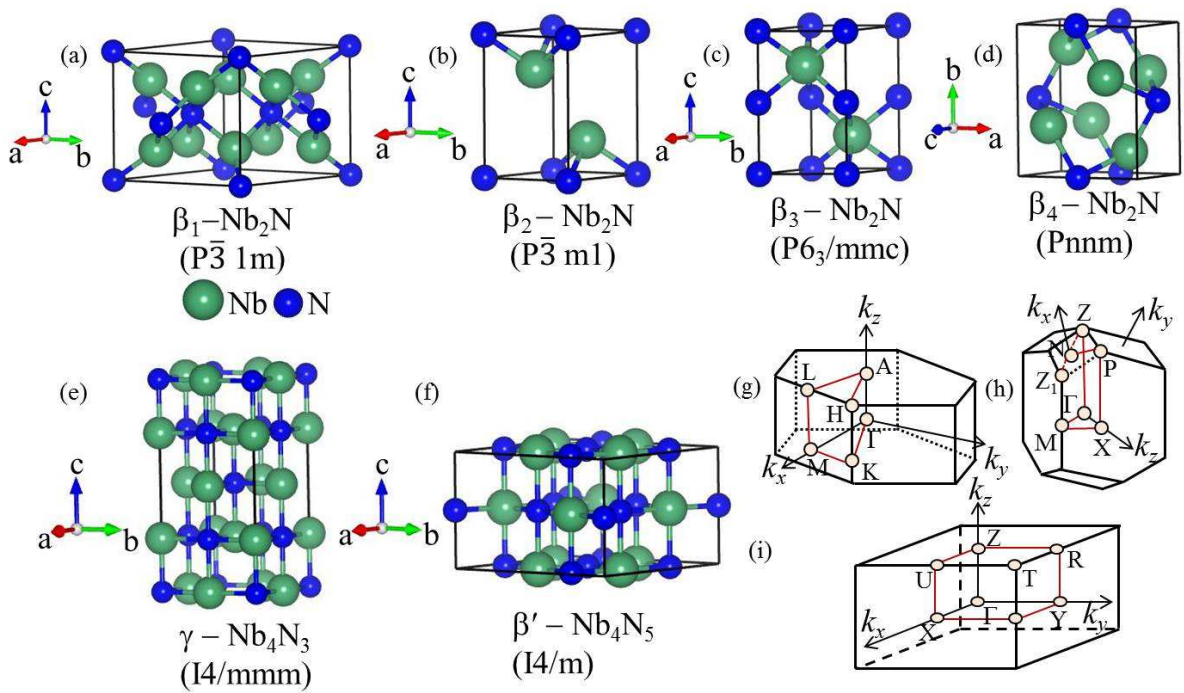}
\caption{Crystal structures of (a) $\beta_1$-Nb$_2$N, (b) $\beta_2$-Nb$_2$N, (c) $\beta_3$-Nb$_2$N,
(d) $\beta_4$-Nb$_2$N, (e) $\gamma$-Nb$_4$N$_3$ and (f) $\beta^\prime$-Nb$_4$N$_5$.
The hexagonal Brillouin zone (BZ) of $\beta_1$-Nb$_2$N, $\beta_2$-Nb$_2$N and $\beta_3$-Nb$_2$N is illustrated in (g),
the tetragonal BZ of $\gamma$-Nb$_4$N$_3$ and $\beta^\prime$-Nb$_4$N$_5$ is shown in (h), and
the orthorhombic BZ of $\beta_4$-Nb$_2$N is plotted in (i).
}
\end{figure*}

\begin{table}[ht]
\caption{Crystal structure, space group, and superconducting transition temperature $T_c$ of some niobium nitrides.}
\begin{tabular}{cccc} \hline \hline
Phase & Structure & Space group & $T_c$ (K) \\ \hline
$\alpha$-NbN & Cubic & Pm$\bar{3}$m & 16\footnotemark[1] \\
$\delta$-NbN & Cubic & Fm$\bar{3}$m & 17.3\footnotemark[2] \\
$\delta^\prime$-NbN & Hexagonal & P6$_3$/mmc  &  $<1.77$\footnotemark[3] \\
$\varepsilon$-NbN & Hexagonal & P6$_3$/mmc & 11.6\footnotemark[4], $<1.77$\footnotemark[3] \\
WC-NbN & Hexagonal & P$\bar{6}$m2 & \\
$\beta_1$-Nb$_2$N & Trigonal  & P$\bar{3}$1m & 8.6-12.1\footnotemark[5], $<1$\footnotemark[6], 4.74 \footnotemark[7] \\
$\beta_2$-Nb$_2$N & Trigonal  & P$\bar{3}$m1 &                                                \\
$\beta_3$-Nb$_2$N & Hexagonal &  P6$_3$/mmc  &                                             \\
$\beta_4$-Nb$_2$N & Orthorhombic & Pnnm  & \\
$\gamma$-Nb$_4$N$_3$& Tetragonal & I4/mmm & 7.8-12.2\footnotemark[8] \\
$\beta^\prime$-Nb$_4$N$_5$ & Tetragonal & I4/m & 10\footnotemark[9], 8-16\footnotemark[3]  \\ \hline \hline
\end{tabular}\\
\footnotemark[1]{References [\onlinecite{Brauer, GBrauer}] (experiment);}
\footnotemark[2]{Reference [\onlinecite{Gold}] (experiment);}
\footnotemark[3]{Reference [\onlinecite{Oya}] (experiment);}
\footnotemark[4]{Reference [\onlinecite{Zou}] (experiment);}
\footnotemark[5]{Reference [\onlinecite{Gavaler}] (experiment);}
\footnotemark[6]{Reference [\onlinecite{Jena}] (experiment);}
\footnotemark[7]{Reference [\onlinecite{Kalal}] (experiment);}
\footnotemark[8]{Reference [\onlinecite{Kaiser}] (experiment);}
\footnotemark[9]{Reference [\onlinecite{Merch}] (experiment);}
\end{table}

The rest of this paper is organized as follows. In section II, we introduce the crystal structures
of the considered nitrides, theory of superconductivity,  {\it ab initio} calculation methods and computational details used
in the present study. In section III, the calculated physical properties of the niobium nitrides are presented.
In particular, the theoretical elastic constants, moduli and hardness of the nitrides
are reported in Section III A. In section III B, the calculated electronic band structures are presented and
Dirac nodal points are identified. In section III C, the calculated phonon dispersion relations
as well as the contributions from the lattice vibrations and conduction electrons to the specific heat
and Debye temperatures are presented. Finally, the calculated electron-phonon coupling strengths
and estimated superconducting transition temperatures are reported in section III D.
In section IV, we summarize the conclusions drawn from this work.

\section{CRYSTAL STRUCTURES AND COMPUTATIONAL METHODS}
The crystal structures and the corresponding Brillouin zones of all the considered niobium nitrides
are shown in Fig. 1. Four crystalline structures have been reported for $\beta$-Nb$_2$N, namely,
trigonal P$\bar{3}$1m (No.162) \cite{Norlund} ($\beta_1$-Nb$_2$N) and P$\bar{3}$m1 (No.164) \cite{Savage}
($\beta_2$-Nb$_2$N), hexagonal P6$_3$mmc (No.194) \cite{Tera} ($\beta_3$-Nb$_2$N) and
orthorhombic Pnnm (No. 58)\cite{Zha} ($\beta_4$-Nb$_2$N).
The crystal structure of $\beta_1$-Nb$_2$N contains three formula units (f.u.) per unit cell \cite{Norlund}.
Nb occupies the wyckoff site 6$k$ ($\frac{1}{3}$, 0, $\frac{1}{4}$), N is at 1$a$ (0, 0, 0) and 2$d$ ($\frac{1}{3}$,
$\frac{2}{3}$, $\frac{1}{2}$). In $\beta_2$-Nb$_2$N, Nb is at 2$c$ ($\frac{1}{3}$, $\frac{2}{3}$, $\frac{1}{4}$)
and N at 2$a$ (0, 0, 0) whereas
in $\beta_3$-Nb$_2$N, Nb occupies 2$c$ ($\frac{1}{3}$, $\frac{2}{3}$, $\frac{1}{4}$) and N is at 2$a$ (0, 0, 0).
The crystal structure of $\beta_4$-Nb$_2$N has two f.u.
per unit cell \cite{Zha}. Nb occupies 4$g$ (0.2572, 0.3390, 0) and N 2$d$ ($\frac{1}{2}$, 0, 0).
Both $\gamma$-Nb$_4$N$_3$ and $\beta^\prime$-Nb$_4$N$_5$ crystallize in the tetragonal structure
with space group I4/mmm (No. 139) \cite{Pearson} and I4/m (No. 7) \cite{Terao}, respectively.
The unit cell of $\gamma$-Nb$_4$N$_3$ contains two f.u. with Nb at 4$c$ (0, $\frac{1}{2}$, 0)
and 4$e$ (0, 0, 0.2521) and N atoms at 2$a$ (0, 0, 0) and 4$d$ (0, $\frac{1}{2}$, $\frac{1}{4}$).
The unit cell of $\beta^\prime$-Nb$_4$N$_5$ has two f.u. with Nb at 8$h$ (0.4, 0.2, 0) and N
at 2$b$ (0, 0, $\frac{1}{2}$) and 8$h$ (0.1, 0.3, 0). It is worth to mention that all the
crystal structures possess inversion ($\mathcal{P}$) symmetry.

The {\it ab initio} structural optimizations, elastic constants, electronic band structures
and density of states (DOS) calculations are based on density functional theory (DFT)
with the generalized gradient approximation (GGA) \cite{PBE}.
The calculations are performed by using the accurate projector-augmented wave
method \cite{GKresse,PEBlochl,Furthmuller}, as implemented in the Vienna \textit{Ab initio} Simulation
Package (VASP). For the Brillouin zone integration, the tetrahedron method is used with $\Gamma$-centered $k$-point
meshes of 8$\times$8$\times$10, 8$\times$8$\times$6, 8$\times$8$\times$6, 8$\times$6$\times$10,
8$\times$8$\times$4 and 8$\times$8$\times$10, respectively, for $\beta_1$-Nb$_2$N,
$\beta_2$-Nb$_2$N, $\beta_3$-Nb$_2$N, $\beta_4$-Nb$_2$N, $\gamma$-Nb$_4$N$_3$
and $\beta^\prime$-Nb$_4$N$_5$. A large plane-wave cut-off energy of 500 eV is used throughout.
The DOS are calculated by using denser $k$-point meshes of 16 $\times$ 16 $\times$ 20
for $\beta_1$-Nb$_2$N, 16 $\times$ 16 $\times$ 12 for $\beta_2$-Nb$_2$N and $\beta_3$-Nb$_2$N,
and 16 $\times$ 12 $\times$ 20 for $\beta_4$-Nb$_2$N, 16 $\times$ 16 $\times$ 8 for $\gamma$-Nb$_4$N$_3$
and 16 $\times$ 16 $\times$ 20 for $\delta^\prime$-NbN. In the crystal structure optimizations,
the structures are relaxed until the atomic forces are less than 0.0001 eV/\AA.
A small total energy convergence criterion of 10$^{-8}$ eV is used for all the calculations.
The calculated lattice constants and total energies for all the considered structures
are listed in Table II.
We notice that the calculated lattice constants
of all the structures are in good accord with the available experiment
data~\cite{Norlund,Zha,Savage,Tera,Pearson,Terao} and previous
theoretical calculations based on GGA. \cite{Daul, Chihi, Riyan}
Among the four structures of $\beta$-Nb$_2$N, $\beta_1$-Nb$_2$N is found to be
the ground state structure with $\beta_4$-Nb$_2$N, $\beta_2$-Nb$_2$N, $\beta_3$-Nb$_2$N structures being,
respectively, 0.062 eV/f.u., 0.305 eV/f.u. and 0.381 eV/f.u. higher in total energy.

The elastic constants of the niobium nitrides are calculated by using the linear-response
stress-strain method, as implemented in the VASP code\cite{Page}. Under a small strain ($\varepsilon_{kl}$),
according to Hooke's law, the corresponding stress ($\sigma_{ij}$) can be written as $\sigma_{ij}$ = $C_{ijkl}\varepsilon_{kl}$,
where $C_{ijkl}$ is the elastic stiffness tensor that consists of the elastic constants of the crystal.
The total number of elastic constants depends on the crystal symmetry. The calculated nonzero elastic constants
for all the considered structures are listed in Table III.
For the hexagonal and trigonal crystals, bulk modulus $B$ and shear modulus $G$ are given
by $ B = \frac{2}{9} (C_{11} + C_{12} + 2C_{13} + \frac{1}{2}C_{33})$ and $G = \frac{1}{30}(12C_{44} +
7C_{11} - 5C_{12} + 2C_{33} - 4C_{13})$. For the tetragonal structures,
$B = \frac{1}{9}$\{2($C_{11}$ + $C_{12}$) + $C_{33}$ + 4$C_{13}$\} and
$G = \frac{1}{30}$\{4$C_{11}$ - 2$C_{12}$ - 4$C_{13}$ + 2$C_{33}$ + 12$C_{44}$ + 6$C_{66}$\}.
In the orthorhombic crystals, $B = \frac{1}{9}$\{$C_{11}$ + $C_{22}$ + $C_{33}$ + 2$C_{12}$ + 2$C_{13}$ + 2$C_{23}$\} and
$G = \frac{1}{15}$\{$C_{11}$ + $C_{22}$ + $C_{33}$ - ($C_{12}$ + $C_{13}$ + $C_{23}$)\} + $\frac{3}{15}$\{$C_{44}$ + $C_{55}$ + $C_{66}$\}.
The Young's modulus \textit{Y} and Poisson's ratio are related to \textit{B} and \textit{G} by
$Y = 9BG/(3B + G)$ and $\nu = (3B - 2G)/2(3B + G)$. The hardness \textit{H}
can be estimated by $H =0.1769G - 2.899$.~\cite{Khenata}

For superconductors with the dominant electron-phonon interaction,
the superconductiviting properties can be analyzed through calculating the Eliashberg spectral
function $\alpha^2F(\omega)$. Hence, we calculate the phonon dispersion relations, phonon DOS
and electron-phonon interactions using the {\it ab initio} density functional perturbation theory (DFPT),
\cite{SBaroni} as implemented in the Quantum Espresso code. \cite{PGianozzi}
The calculations are performed with the scalar-relativistic optimized norm-conserving Vanderbilt
pseudopotentials. \cite{Haman, JPPerdew} The plane wave cut-off energy is set to 42 Ry and the
electronic charge density is expanded up to 168 Ry. A Gaussian broadening of 0.02 Ry is used
for all the calculations. All the phonon and electron-phonon coupling calculations are perfomed
with a $q$-grids of 4$\times$4$\times$5, 4$\times$4$\times$3, 4$\times$4$\times$3, 4$\times$3$\times$5,
4$\times$4$\times$2, and 3$\times$3 $\times$ 4 for $\beta_1$-Nb$_2$N, $\beta_2$-Nb$_2$N, $\beta_3$-Nb$_2$N,
$\beta_4$-Nb$_2$N, $\gamma$-Nb$_4$N$_3$ and $\beta^\prime$-Nb$_4$N$_5$, respectively.

The strength of the electron-phonon coupling in a crystal is measured by the electron-phonon coupling
constant ($\lambda$), which can be extracted from the Eliashberg spectral function [$\alpha^2F(\omega)$]
via the Allen-Dynes formula ~\cite{McMillan, AllenDynes}
\begin{equation}
\lambda = 2 \int \frac{\alpha^2F(\omega)}{\omega}d\omega.
\end{equation}
The Eliashberg spectral function is given by
\begin{equation}
\alpha^2F(\omega)=\frac{1}{2\pi N(E_F)}\sum_{qj} \frac{\gamma_{qj}}{\omega_{qj}}\delta(\hbar\omega-\hbar\omega_{qj}),
\end{equation}
where $N(E_F)$ is the electronic DOS at the Fermi level ($E_F$), $\gamma_{qj}$
is the phonon linewidth due to electron-phonon scattering, $\omega_{qj}$ is the phonon frequency
of branch index $j$ at wave vector $q$.
Using the calculated $\lambda$, one can estimate the superconducting transition temperature $T_c$
via McMillan-Allen-Dynes formula \cite{McMillan, AllenDynes}
\begin{equation}
T_c = \frac{\omega_{log}}{1.2} \textrm{exp}\Big[\frac{-1.04{(1+\lambda)}}{\lambda-\mu^*(1+0.62\lambda)}\Big],
\end{equation}
where $\omega_{log}$ is logarithmically averaged phonon frequency and $\mu^*$ is the averaged
screened electron-electron interaction.

\begin{table}[ht]
\caption{Theoretical lattice constants ($a, b, c, c/a$), volume ($V$) and total energy ($E_t$)
of all the studied niobium nitrides, compared with the available experimental data (Expt).
}
\begin{tabular}{cccccccc} \hline \hline
Phase                      & $a$ (\AA)&$b$ (\AA)&$c$ (\AA)&$c/a$& V (\AA$^3$/at) & $E_t$ (eV/at)& \\ \hline
$\beta_1$-Nb$_2$N          & 5.341 &     & 5.009 & 0.937 & 13.75 & -10.3743   \\
Expt\footnotemark[1]       & 5.267 &     & 4.987 & 0.946 &       &      \\
$\beta_2$-Nb$_2$N          & 3.157 &     & 4.858 & 1.538 & 13.98 & -10.2725    \\
Expt\footnotemark[2]       & 3.058 &     & 4.961 & 1.622 &       &      \\
$\beta_3$-Nb$_2$N          & 2.999 &     & 5.605 & 1.868 & 14.56 & -10.2470    \\
Expt\footnotemark[3]       & 3.055 &     & 4.994 & 1.634 &       &     \\
$\beta_4$-Nb$_2$N          & 4.931 &5.455& 3.066 & 0.562 & 13.75 & -10.3536    \\
$\gamma$-Nb$_4$N$_3$       & 4.427 &     & 8.707 & 1.966 & 12.19 & -10.2478     \\
 Expt\footnotemark[4]      & 4.382 &     & 8.632 & 1.969 &       &     \\
$\beta^\prime$-Nb$_4$N$_5$ & 6.933 &     & 4.324 & 0.624 & 11.55 & -10.0132     \\
 Expt\footnotemark[5]      & 6.873 &     & 4.298 & 0.625 &       &   \\  \hline \hline
\end{tabular}\\
\footnotemark[1]{References [\onlinecite{Norlund}] (experiment);}
\footnotemark[2]{Reference [\onlinecite{Savage}] (experiment);}
\footnotemark[3]{Reference [\onlinecite{Tera}] (experiment);}
\footnotemark[4]{Reference [\onlinecite{Pearson}] (experiment);}
\footnotemark[5]{Reference [\onlinecite{Terao}] (experiment).}
\end{table}

\section{RESULTS AND DISCUSSION}
\subsection{Mechanical properties}

\begin{table*}[ht]
\caption{Calculated elastic constants ($C_{ij}$), bulk modulus ($B$), shear modulus ($G$), Young's modulus ($Y$),
hardness ($H$), Poisson's ratio ($\nu$) and $B$/$G$ ratio of all the considered niobium nitrides.
For comparison, the previous theoretical elastic constants of cubic $\delta$-NbN~\cite{KR} and the experimental elastic constants
of the hard sapphire ($\alpha$-Al$_2$O$_3$)~\cite{Hovis,Teter} are also listed. Also listed are the experimental
$H$ values for $\beta_3$-Nb$_2$N \cite{Sanjines,Zhengbing}. The quantities $C_{ij}$, $B$, $G$, $Y$ and $H$ are in units of GPa.}
\begin{tabular}{cccccccccc} \hline \hline
 & $\beta_1$-Nb$_2$N& $\beta_2$-Nb$_2$N &$\beta_3$-Nb$_2$N  & $\beta_4$-Nb$_2$N & $\gamma$-Nb$_4$N$_3$ & $\beta^\prime$-Nb$_4$N$_5$
 & $\delta$-NbN\footnotemark[1] & $\alpha$-Al$_2$O$_3$ \\ \hline
$C_{11}$ & 417 & 402 & 555 & 399 & 597 & 508 & 692 (608\footnotemark[2])& 497\footnotemark[3] \\
$C_{12}$ & 167 & 102 & 229 & 177 & 89  & 133 & 145 (134\footnotemark[2])& 163\footnotemark[3] \\
$C_{13}$ & 184 & 173 & 168 & 181 & 187 & 134 &     & 116\footnotemark[3] \\
$C_{14}$ &     &    &     &     &     &     &     &  22\footnotemark[3] \\
$C_{23}$ &     &    &     & 158 &     &     &     &  \\
$C_{22}$ &     &    &     & 420 &     &     &     &  \\
$C_{33}$ & 421 & 386 & 619 & 406 & 392 & 643 &    & 501\footnotemark[3] \\
$C_{44}$ & 125 & 150 & 163 & 140 & 107 & 154 & 65 (117\footnotemark[2]) & 147\footnotemark[3] \\
$C_{55}$ &     &    &     & 126 &     &     &    &  \\
$C_{66}$ &     &    &     & 116 & 91  &125  &    &  \\
$B$      & 258 & 232 & 318 & 257 & 280 &273  & 327 (292\footnotemark[2]) & 246\footnotemark[4] \\
$G$      & 123 &  140 & 175 & 124 & 136 & 170 & 148 (165\footnotemark[2]) & 162\footnotemark[4] \\
$Y$      & 318 &  348 & 443 & 319 &351   & 422 & 385 & \\
$H$      & 18.8& 21.9 & 28.1& 19.1&21.2  & 27.2 & 23 & 22\footnotemark[4] \\
         &     &      & (35\footnotemark[5], 30.9\footnotemark[6]) &  & & &  & & \\
$\nu$    & 0.29 & 0.25& 0.27 &0.28 & 0.29 & 0.24 & & & \\
$B$/$G$  & 2.10 &1.66 & 1.82 & 2.07 & 2.06 & 1.61 &  & & \\  \hline \hline
\end{tabular}\\
\footnotemark[1]{Reference [\onlinecite{KR}] ({\it ab initio} calculation);}
\footnotemark[2]{Reference [\onlinecite{Hemley}] (experiment);}
\footnotemark[3]{Reference [\onlinecite{Hovis}] (experiment);}
\footnotemark[4]{Reference [\onlinecite{Teter}] (experiment);}
\footnotemark[5]{Reference [\onlinecite{Sanjines}] (experiment);}
\footnotemark[6]{Reference [\onlinecite{Zhengbing}] (experiment).}
\end{table*}

Elastic constants of a solid provide insight into mechanical stability and bonding characteristics of the material.
In Table III, we list the calculated elastic constants of all the considered niobium nitrides along with
the reported values of well-known hard material sapphire ($\alpha$-Al$_2$O$_3$,
space group R$\bar{3}$c)\cite{Hovis,Teter}.
Table III shows that all the elastic constants are positive, thereby indicating that
all the considered nitride structures are mechanically stable against the corresponding
specific deformations \cite{Born}.
Table III also shows that for $\beta_1$-Nb$_2$N, $\beta_3$-Nb$_2$N, $\beta_4$-Nb$_2$N
and $\beta^\prime$-Nb$_4$N$_5$, $C_{33}$ is larger than $C_{11}$,
indicating that the materials are harder to compress along the $c$-axis while it is softer for
both $\beta_2$-Nb$_2$N and $\gamma$-Nb$_4$N$_3$. The calculated elastic moduli suggest
that all the nitrides are hard materials.
In particular, the calculated bulk modulus ($B$) of the niobium nitrides is either comparable
to or larger than that of hard sapphire \cite{Teter}.
For example, for $\beta_3$-Nb$_2$N, the calculated $B$  value is about 30\%
larger than the corresponding value of sapphire \cite{Teter}.
Interestingly, when compared to the
niobium mononitride structures \cite{KR}, e.g., $\delta$-NbN ($B$ = 327 GPa),
both Nb-rich $\beta$-Nb$_2$N and
$\gamma$-Nb$_4$N$_3$ as well as N-rich
$\beta^\prime$-Nb$_4$N$_5$ possess up to about 20 \% lower $B$ values. This
indicates that both Nb-rich and N-rich niobium nitrides are softer materials compared to the niobium mononitride.
Furthermore, the $B$ of all the nitride structures is almost twice that of
the shear modulus $G$, suggesting that $G$ is the limiting parameter for the mechnaical stability.

Young's modulus ($Y$) of a solid is the ratio of linear stress to strain and tells us about the stiffness
of the material. The calculated $Y$ of $\beta_3$-Nb$_2$N and $\beta^\prime$-Nb$_4$N$_5$
is about 25 \% larger than that of the other nitrides, indicating their higher stiffness.
According to Pugh's criteria~\cite{Pugh}, the value of $B$/$G$ greater than (less than) 1.75 would inidcate
ductile (brittle) character of the material. Table III thus shows that $\beta_2$-Nb$_2$N
($B$/$G$ = 1.66) and $\beta^\prime$-Nb$_4$N$_5$ ($B$/$G$ = 1.61) are brittle materials
while the rest ($B$/$G$ $>$ 1.75) are ductile materials. In particular,
$\beta_1$-Nb$_2$N ($B$/$G$ = 2.10) is more ductile than all the other nitrides.
Poisson's ratio $\nu$ measures the stability of a material against the shear strain.
Among the studied nitrides, $\beta^\prime$-Nb$_4$N$_5$ has the smallest value ($\nu$ = 0.24)
indicating that it is relatively stable against shear strain compared to the other nitrides.
Hardness ($H$) is an important elastic property which is responsible for wear behaviour
of materials \cite{Khenata}. It is clear from Table III that $\beta_3$-Nb$_2$N
has the strongest hardness followed by $\beta^\prime$-Nb$_4$N$_5$, $\beta_2$-Nb$_2$N,
$\gamma$-Nb$_4$N$_3$, $\beta_4$-Nb$_2$N, and $\beta_1$-Nb$_2$N.
The calculated hardness value of $\beta_3$-Nb$_2$N (28.1 GPa)
is close to the experimental values of 35 GPa\cite{Sanjines} and 30.9 GPa\cite{Zhengbing}.
Importantly, Table III shows that the hardness $H$ of the nitrides $\beta_1$-Nb$_2$N, $\beta_2$-Nb$_2$N,
$\beta_4$-Nb$_2$N and $\gamma$-Nb$_4$N$_3$ is close to that of hard sapphire \cite{Teter}.
Both $\beta_3$-Nb$_2$N and $\beta^\prime$-Nb$_4$N$_5$
are harder than sapphire because of almost 40 \% larger $H$ values (Table III).

\begin{figure}
\centering
\includegraphics[width=80mm]{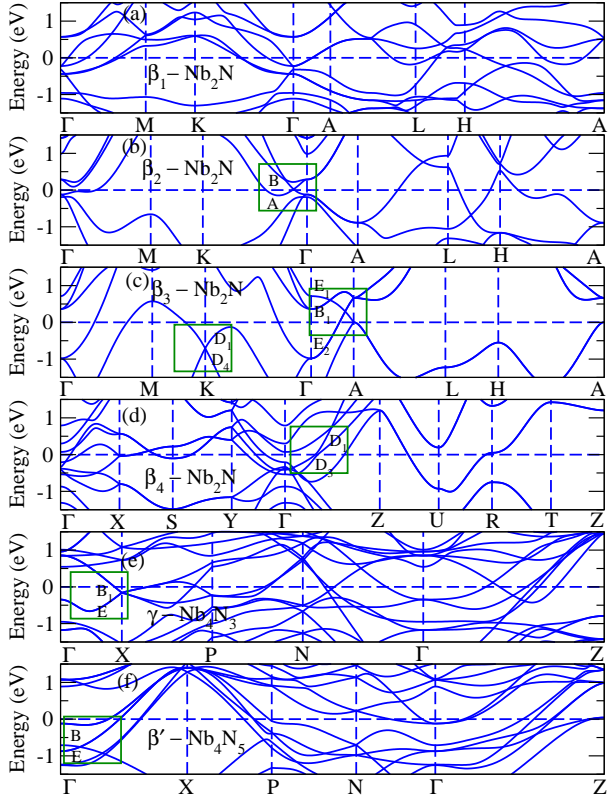}
\caption{Electronic band structures of (a) $\beta_1$-Nb$_2$N, (b) $\beta_2$-Nb$_2$N, (c) $\beta_3$-Nb$_2$N,
(d) $\beta_4$-Nb$_2$N, (e) $\gamma$-Nb$_4$N$_3$ and (f) $\beta^\prime$-Nb$_4$N$_5$
calculated without SOC. The green boxes indicate the band crossings with symmetries of the bands labeled.}
\end{figure}

\subsection{Band structure and Dirac nodal points}

\begin{figure}
\centering
\includegraphics[width=80mm]{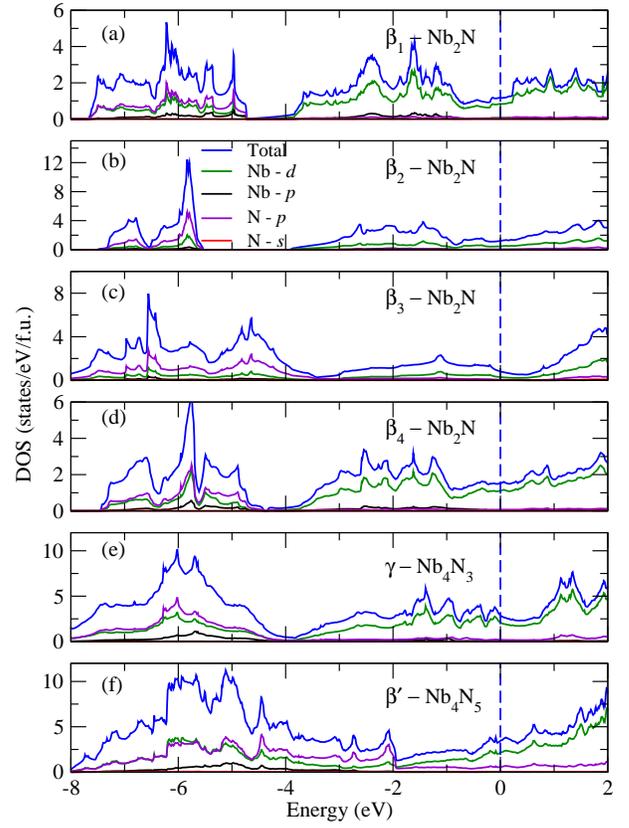}
\caption{Total and orbital-decomposed density of states (DOS) of (a) $\beta_1$-Nb$_2$N,
(b) $\beta_2$-Nb$_2$N, (c) $\beta_3$-Nb$_2$N, (d) $\beta_4$-Nb$_2$N, (e) $\gamma$-Nb$_4$N$_3$
and (f) $\beta^\prime$-Nb$_4$N$_5$.}
\end{figure}

The energy bands and DOS spectra calculated without including the spin-orbit coupling (SOC),
of all the studied structures are displayed in Figs. 2 and 3, respectively.
Figures 2 and 3 show that all the considered Nb nitrides are metallic with many bands crossing
$E_F$ and also have relatively large DOS at $E_F$ (see Table IV).
Interestingly, Figs. 3(a) and 3(d) show that the DOS spectra of $\beta_1$-Nb$_2$N and $\beta_4$-Nb$_2$N
are very similar, although their structures are quite different (Table I and Fig. 1).
This implies that they have similar bonding characterstics.
Indeed, this explain why their elastic moduli ($B$, $G$, $Y$ and $H$) are very similar (see Table III).
In particular, in the lower valence band region below -4.0 eV, Nb $d$ DOS and N $p$ DOS
spectra in both cases have nearly the same magnitudes, indicating a strong covalent bonding in
these two structures [Figs. 3(a) and 3(d)]. This is also the case for $\gamma$-Nb$_4$N$_3$ [Fig. 3(e)]
and $\beta^\prime$-Nb$_4$N$_5$ in the region below -2.0 eV [Fig. 3(f)].
Nevertheless, the weight of Nb $d$ states in $\beta_2$-Nb$_2$N and $\beta_3$-Nb$_2$N becomes
significantly smaller than that of N $p$ states, indicating that the covalency in these nitrides decreases.
This explains that the tetragonal $\gamma$-Nb$_4$N$_3$
and $\beta^\prime$-Nb$_4$N$_5$ has superior mechanical properties compared to $\beta_1$-Nb$_2$N.

On the other hand, Fig. 3 indicates that the upper valence bands and lower conduction bands
from -4.0 to 2.0 eV of $\beta_1$-Nb$_2$N, $\beta_4$-Nb$_2$N and $\gamma$-Nb$_4$N$_3$
are Nb $d$ dominated states. This is also the case for $\beta'$-Nb$_4$N$_5$
from -2.0 to 2.0 eV (see Fig. 3). Thus, the Nb $d$ states are important for governing
the superconducting and other transport properties of these nitrides.
Furthermore, for $\beta_1$-Nb$_2$N, $\beta_2$-Nb$_2$N, and $\beta_4$-Nb$_2$N,
the DOS in the vicinity of the $E_F$ is nearly constant
and takes value of 0.585 states/eV/Nb, 0.705 states/eV/Nb and 0.755 states/eV/Nb at $E_F$,
respectively (see Fig. 3 and Table IV).
In contrast, in $\beta_3$-Nb$_2$N [Fig. 3(c)], $\gamma$-Nb$_4$N$_3$ [Fig. 3(e)] and
$\beta^\prime$-Nb$_4$N$_5$ [Fig. 3(f)], the DOS monotonically decreases with energy and
at $E_F$ has the value of 0.610 states/eV/Nb, 0.698 states/eV/Nb, and 0.813 states/eV/Nb,
respectively.

Interestingly, when the SOC is neglected, the band structure exhibits symmetry protected band crossings
in the vicinity of the Fermi level along $k$-paths K-$\Gamma$,
$\Gamma$-A and $\Gamma$-Z for $\beta_2$-Nb$_2$N, $\beta_3$-Nb$_2$N and $\beta_4$-Nb$_2$N, respectively
(see Fig. 2). Such band crossings also occur
along $\Gamma$-X for $\gamma$-Nb$_4$N$_3$ and $\beta^\prime$-Nb$_4$N$_5$.
For  $\beta_4$-Nb$_2$N, $k$-path $\Gamma$-Z belongs to the $C_{2v}$ point group
and the bands have two different irreducible represnetation (IRs) D$_1$ and D$_3$.
In $\beta_3$-Nb$_2$N, $\Gamma$-A line has the $C_{6v}$
point group symmetry and there exist two band crossings at 1.0 eV below the $E_F$ [
Fig. 2(d)]. At about 0.2 eV above the $E_F$, the band crossing is between a nondegenerate band
with IR B$_1$ and a doubly degenerate band with IR E$_2$. The other band crossing at $\sim$ 0.6 eV
involves two different bands with IRs B$_1$ and E$_1$, respectively. These two band crossings
are protected by the $C_{3z}$ rotational symmetry. Another symmetry protected band crossing
with IRs D$_1$ (A$_1$) and D$_4$ (B$_2$) is visible at the high symmetry $k$-point K which
belongs to the $C_{2v}$ point group. Further, the band structure of $\beta_2$-Nb$_2$N
shows a band crossing along K-$\Gamma$ between the bands with different IRs A and B,
which belong to the $C_2$ point group symmetry and hence are forbidden to mix.
For tetragonal $\gamma$-Nb$_4$N$_3$, the linear band crossing between IRs B$_1$ and
E is located along the $\Gamma$-X direction and it is protected by the $C_{4v}$ point group symmetry.
In $\beta^\prime$-Nb$_4$N$_5$, the bands with IRs B and E cross each other along the $\Gamma$-X path
and are protected by the $\mathcal {C}_{4}$ rotational symmetry of the $C_{4v}$ point group.

Fully relativistic band structures of the nitrides are shown in Fig. 4.
When the SOC is included, significant changes in the band structure occur. Among other things,
the single point group symmetry changes to the double point
group symmetry and hence the IRs of the bands change as well. Importantly,
since the SOC breaks SU(2) symmmetry, some band crossings would become gapped.
For example, the IRs $D_1$ and $D_3$ of the bands crossing along $\Gamma$-Z
in $\beta_4$-Nb$_2$N as well as the IRs A and B of the bands crossing
along $\Gamma$-K direction for $\beta_2$-Nb$_2$N now become
$\Gamma$$_5$ [Figs. 4(d) and 4(j)] and $\Gamma$$_3$ [Figs. 4(b) and 4(g)], respectively.
Both band crossings are now gapped [see Figs. 4(g) and 4(j)].
The band crossing at the M point in $\beta_3$-Nb$_2$N
is now represented by $\Gamma_5$ ($D_5$) [Fig. 4(d)] and it is gapped by $\sim$ 0.05 eV.

\begin{table*}[ht]
\caption{Calculated electron-phonon coupling constant ($\lambda$), logarithmic average phonon frequency
($\omega_{log}$), density of states at the Fermi level [$N(E_F)$], low temperature
specific heat coefficients ($\gamma$ and $\beta$),
Debye temperature ($\Theta_{D}$) and superconducting transition temperature ($T_c$)
of all the studied niobium nitrides.
The smearing parameter ($\sigma$) used is 0.02 Ry.
The screened Coulomb interaction $\mu^*$ is set to 0.10.
Available experimental $T_c$ values are also listed for comparison.}
\begin{ruledtabular}
\begin{tabular}{cccccccc}
Structure      & $\lambda$ & $\omega_{log}$  & $N(E_F)$ &$\gamma$ &$\beta$ &$\Theta_{D}$ &$T_c$  \\
                 &      &            (K)     & (states/eV/Nb) & (mJ/mol-K$^2$)&(mJ/mol-K$^4$)& (K) & (K)  \\ \hline
$\beta_1$-Nb$_2$N& 0.36 & 285 & 0.585 &2.75 &0.460 & 233 & 0.57    \\
Expt             &      &     &              & & & &  8.6-12.1  \footnotemark[1]    \\
$\beta_2$-Nb$_2$N& 0.57 & 306 & 0.705 &3.32 & 0.220 & 298& 6.12      \\
Expt             &      &                &     & & & &  8.6-12.1  \footnotemark[1]    \\
$\beta_3$-Nb$_2$N& 0.47 & 398 & 0.610 &2.87 & 0.163 & 330 & 3.88      \\
Expt             & 0.54 &     &       &     &       & 320\footnotemark[2] & $<$ 1 \footnotemark[3], 4.74  \footnotemark[2]    \\
$\beta_4$-Nb$_2$N& 0.46 & 289 & 0.755 &3.56 &0.270 & 278& 2.64     \\
$\gamma$-Nb$_4$N$_3$&0.68&262 & 0.698 & 6.57 & 0.880 & 249 &8.48      \\
Expt             &   &  & & & & & 7.8-12.2\footnotemark[4], 8-16 \footnotemark[5]  \\
$\beta^\prime$-Nb$_4$N$_5$&0.92&249& 0.873 & 8.22 & 0.824 & 277 & 15.28  \\
Expt             &      &      &     &  & & & 8-16 \footnotemark[5], 10 \footnotemark[6] \\
$\delta$-NbN     & 0.98 & 269 & 0.883  &2.09 &0.518 & 196 & 18.3    \\
\end{tabular}
\end{ruledtabular}
\footnotemark[1]{Reference [\onlinecite{Gavaler}] (experiment);}
\footnotemark[2]{Reference [\onlinecite{Kalal}] (experiment);}
\footnotemark[3]{Reference [\onlinecite{Jena}] (experiment);}
\footnotemark[4]{Reference [\onlinecite{Kaiser}] (experiment);}
\footnotemark[5]{Reference [\onlinecite{Oya}] (experiment);}
\footnotemark[6]{Reference [\onlinecite{Merch}] (experiment).}
\end{table*}

Remarkably, several band crossings remain intact after SOC is turned-on.
These survived band crossings include that along the $\Gamma$-A line in hexagonal $\beta_3$-Nb$_2$N
[Fig. 4(c)], $\Gamma$-X line in tetragonal $\gamma$-Nb$_4$N$_3$ [Fig. 4(e)] and $\beta^\prime$-Nb$_4$N$_5$
[Fig. 4(f)].
The two band crossings along the $\Gamma$-A line in $\beta_2$-Nb$_2$N previously between the bands
with IRs B$_1$ and E$_2$, B$_1$ and E$_1$, are now tranformed
from B$_1$ to $\Gamma$$_7$, E$_1$ and E$_2$ to $\Gamma$$_8$ and $\Gamma$$_9$ [Figs. 4(h) and 4(i)].
Consequently, there are now three band crossings which are protected by the mirror plane
and $\mathcal {C}_{3z}$ rotational symmetry. The band crossings in $\gamma$-Nb$_4$N$_3$
along $\Gamma$-X are represented by $\Gamma$$_6$ and $\Gamma$$_7$ [Figs. 4(k) and 4(l)]
and are protected by the $\mathcal {C}_{4}$ rotational symmetry. Furthermore,
the band crossings along $\Gamma$-X in $\beta^\prime$-Nb$_4$N$_5$ belong to the IRs $\Gamma$$_5$
and $\Gamma$$_7$ [Figs. 4(f) and 4(l)] and protected by the $\mathcal {C}_{4}$ symmetry.
All the other band crossings become gapped out when SOC is included. Overall, there exist ungapped band crossings
in the relativistic band structures along $\Gamma$-A for $\beta_4$-Nb$_2$N [Figs. 4(d), 4(i) and 4(j)]
and along $\Gamma$-X for both $\gamma$-Nb$_4$N$_3$ [Figs. 4(e) and 4(k)] and $\beta^\prime$-Nb$_4$N$_5$
[Figs. 4(f) and 4(l)]. This demonstrates that Nb-rich $\beta_4$-Nb$_2$N, $\gamma$-Nb$_4$N$_3$
and N-rich $\beta^\prime$-Nb$_4$N$_5$ are topological metals. Importantly, all these three structures
have both time-reversal ($\mathcal{T}$) symmetry and inversion ($\mathcal{P}$) symmetry
and hence each energy band is two fold degenerate. Therefore, the band crossings are
four fold Dirac points (DP). In particular, the DPs in Nb-rich $\beta_3$-Nb$_2$N
and $\gamma$-Nb$_4$N$_3$ are conventional type I whereas in N-rich $\beta^\prime$-Nb$_4$N$_5$,
the DPs are of type II because the slopes of the two crossing bands have the same sign.

\begin{figure*}
\centering
\includegraphics[width=160mm]{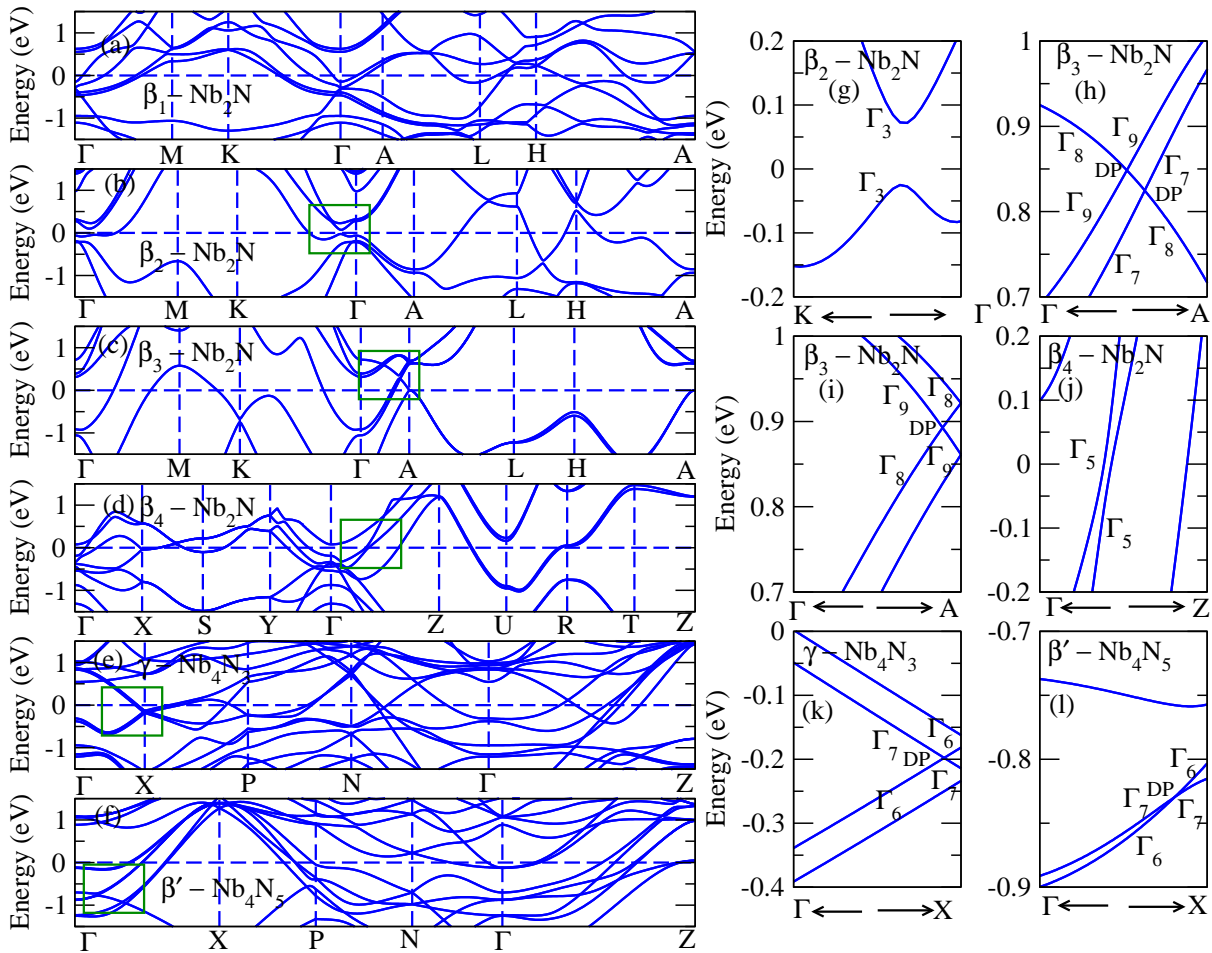}
\caption{Relativistic electronic band structures of (a) $\beta_1$-Nb$_2$N, (b) $\beta_2$-Nb$_2$N,
(c) $\beta_3$-Nb$_2$N, (d) $\beta_4$-Nb$_2$N, (e) $\gamma$-Nb$_4$N$_3$ and (f) $\beta^\prime$-Nb$_4$N$_5$.
The right side panels (g), (h), (i)+(j), (k) and (l) correspond to the zoom-in plots
of the band crossings in the green boxes in (b), (c), (d), (e) and (f), respectively.
DP in (i), (j), (k) and (l) represents the Dirac point. The irreducible representations (IRs)
of the band crossings in (g), (h), (i)+(j), (k) and (l) are also shown.}
\end{figure*}

\subsection{Lattice dynamics and specific heat}
The calculated phonon dispersion relations and phonon DOS spectra of all the considered
niobium nitrides are presented in Fig. 5 and 6.
First, the absence of any imaginary frequencies in the phonon dispersion relations
throughout the Brillouin zone shows the dynamical stability of the niobium nitride structures,
even although some of them are not the ground state structures (Table II). There are
no experimental data available on the phonon dispersion relations of the considered nitrides.
Second, Figs. 5 and 6 indicate that the phonon
dispersions exhibit a large gap between the Nb atom dominated low-energy modes
and the N atom dominated  high-energy modes.
This is due to the large mass difference between the light N atoms and the heavier Nb atoms.
Furthermore, a significant mixing of low-lying optical modes with the acoustic modes
exists, suggesting that a strong bonding between the Nb and the N atoms.

\begin{figure}
\centering
\includegraphics[width=80mm]{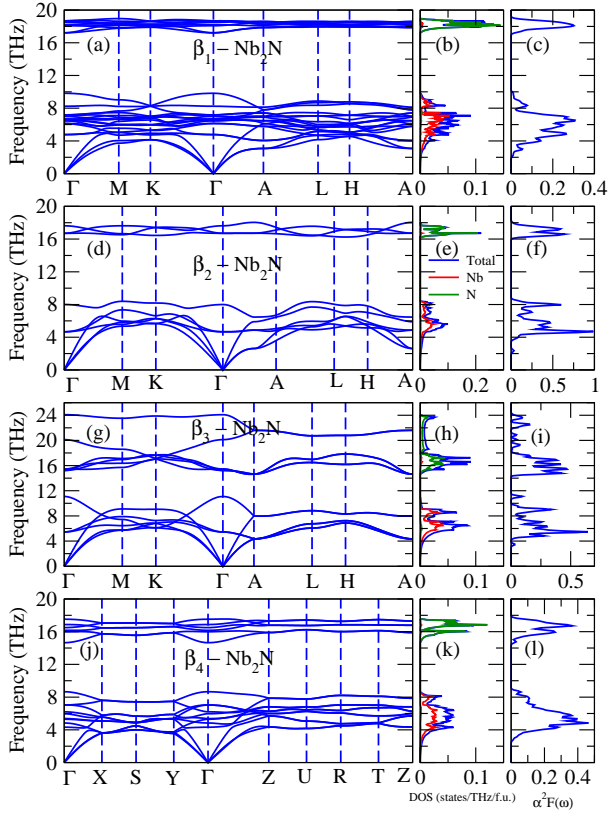}
\caption{Phonon dispersion relations, phonon DOS and Eliashberg function $\alpha^2F(\omega)$
of $\beta_1$-Nb$_2$N (a-c), $\beta_2$-Nb$_2$N (d-f), $\beta_3$-Nb$_2$N (g-i)
and $\beta_4$-Nb$_2$N (j-l).}
\end{figure}

\begin{figure}
\centering
\includegraphics[width=80mm]{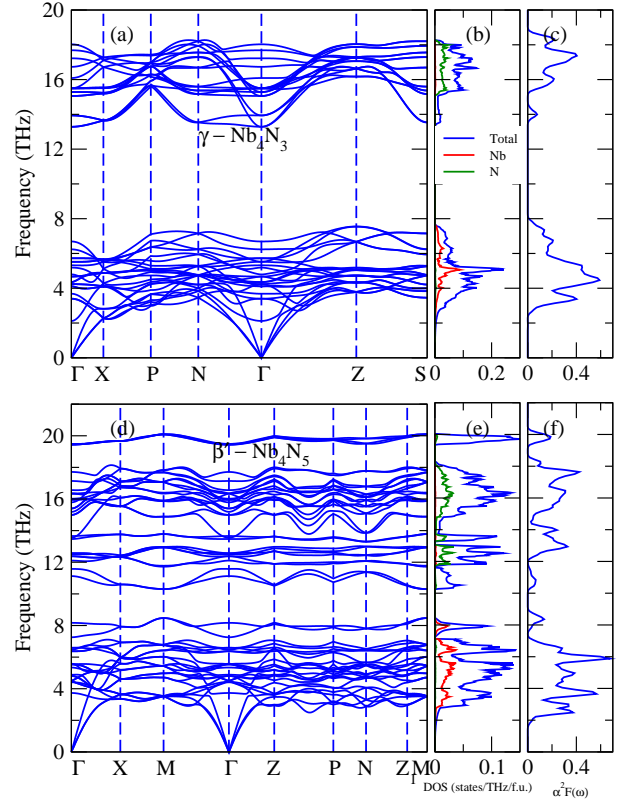}
\caption{Phonon dispersion, phonon DOS and Eliashberg function $\alpha^2F(\omega)$ of $\gamma$-Nb$_4$N$_3$ (a-c)
and $\beta^\prime$-Nb$_4$N$_5$ (d-f).}
\end{figure}

The calculated phonon DOS is used to obtian the specific heat [$C_v(T)$]
with the formula\cite{Kittel}
\begin{equation}
C_v(T) = \gamma T + \int d\omega \frac{(\hbar\omega)^2}{k_B T^2} \frac{g(\omega) e^{\hbar\omega/k_BT}}{(e^{(\hbar\omega/k_BT)}-1)^2} = \gamma T + \beta T^3,
\end{equation}
where the first and second terms are, respectively, the electron and phonon contributions to the specific heat.
Here $\gamma = \frac{\pi^2}{3}k_B^2N(E_F)$ is Sommerfeld coefficient\cite{Sommerfeld} which is proportional to the electron DOS
at $E_F$, $k_B$ is Boltzmann constant and $g(\omega)$ is phonon DOS.
To estimate the coefficient $\beta$ of the phonon contribution at low temperatures,
we first calculate $C_v(T)$ as a function of temperature
between 4 K and 9 K. The calculated $C_v(T)$ is then plotted as $\frac{C_v}{T}$
vs $T^2$ and fitted to $\frac{C_v}{T}$ = $\gamma$ + $\beta$$T^2$.
The resulting values of $\gamma$ and $\beta$ for the considered niobium nitrides are listed in Table IV.
Since $\gamma$ is proportional to $N(E_F)$, $\beta_1$-Nb$_2$N and $\beta^\prime$-Nb$_4$N$_5$ possess
the lowest and highest values of $\gamma$ among the niobium nitrides, respectively.
There are no experimental $\gamma$ data available to compare with.
However, if we incorporate the electron-phonon coupling ($\lambda$),
the electron specific heat $\gamma$ takes the form of $\frac{\gamma_{exp}}{\gamma_{th}}$ = 1 + $\lambda$.
Thus, we can expect that $\frac{\gamma_{exp}}{\gamma_{th}}$ $>$ 1. Finally, the values of $\beta$
are used to calculate the Debye temperature $\Theta_{D}$ by using the relation\cite{Kittel}
$\Theta_{D}$ = ($\frac{12\pi^4N_Ank_B}{5\beta}$)$^\frac{1}{3}$, where $N_A$ is the Avagadro's number
and $n$ is the number of atoms per formula unit. In Table IV, we list the calculated values of $\Theta_{D}$
for the niobium nitride structures. Among the niobium nitrides, $\beta_4$-Nb$_2$N possess the largest $\Theta_{D}$,
which is expected beacuse of its smallest value of $\beta$.
Only the experimental $\Theta_{D}$ value for $\beta_4$-Nb$_2$N has been reported,
which agrees well with the calculated $\Theta_{D}$ value (Table IV).

\subsection{Electron-phonon coupling and superconductivity}
We display in Figs. 5 and 6 the calculated $\alpha^2F(\omega)$ of the studied niobium nitrides.
Equation (2) indicates that $\alpha^2F(\omega)$ is essentially the phonon DOS spectrum modulated by the
electron-phonon interaction matrix element $\gamma_{qj}$ divided by the phonon frequency $\omega_{qj}$.
As a result, the $\alpha^2F(\omega)$ spectrum
for each structure roughly follows the corresponding phonon DOS spectrum (see Figs. 5 and 6).
Therefore, the contribution from the acoustic and low energy
lying optical phonon bands to the $\alpha^2F(\omega)$ may become dominant. This is evident by the existance of
large peaks in the $\alpha^2F(\omega)$ spectrum. Interestingly, among the $\beta$-Nb$_2$N structures,
the magnitude of $\alpha^2F(\omega)$ is highest in the $\beta_3$-Nb$_2$N [Figs. 5(i)] and lowest
for $\beta_1$-Nb$_2$N [5(c)]. For tetragonal structures, $\alpha^2F(\omega)$ of $\beta^\prime$-Nb$_4$N$_5$
shows the larger peaks than for $\gamma$-Nb$_4$N$_3$ [Figs. 6(f) and 6(c)]. Overall, the magnitude
of $\alpha^2F(\omega)$ is highest for $\beta^\prime$-Nb$_4$N$_5$ and lowest for $\beta_1$-Nb$_2$N.
Note that the strength of the electron-phonon coupling ($\lambda$)
is given by an integral of Eliashberg function $\alpha^2F(\omega)$ divided by phonon frequency $\omega$ over
the entire phonon frequency range (Eq. 1).
This results in the lowest value of $\lambda$ (0.36) for $\beta_1$-Nb$_2$N and highest $\lambda$ (0.92)
for $\beta^\prime$-Nb$_4$N$_5$. In Table IV, we list the calculated $\lambda$ values for all the
niobium nitrides. Clearly, the $\lambda$ value (0.92) of $\beta^\prime$-Nb$_4$N$_5$ is much larger
than that of $\gamma$-Nb$_4$N$_3$ (0.68), $\beta_3$-Nb$_2$N (0.47), $\beta_2$-Nb$_2$N (0.57)
and $\beta_1$-Nb$_2$N (0.36).

As mentioned already in section I, the experimental studies \cite{Gavaler,Skokan,Gajar,Jena,Kalal}
on the superconducting properties
of $\beta$-Nb$_2$N so far have reported conflicting results.
Gavaler {\it et al.}\cite{Gavaler} reported the formation
of hexagonal $\beta$-Nb$_2$N in a thin film using the XRD data with some additional peaks
which were not indexed, and also a $T_c$ of 8.6 K in the thin films. In addition,
they also reported another film which has mixed phases of hexagonal $\beta$-Nb$_2$N and cubic-NbN with
a $T_c$ of 12.1 K [1]. Skokan {\it et al.}\cite{Skokan} reported that the thin films of mixed phases
of cubic-NbN and hexagonal $\beta$-Nb$_2$N exhibit two step resistance drop at 9 K and at 2 K.
Gajar {\it et al.} \cite{Gajar} reported  the transformation of Nb into hexagonal $\beta$-Nb$_2$N
which becomes superconducting below 1 K only.
However, Kalal {\it et al.}\cite{Kalal} recently reported that the hexagonal $\beta$-Nb$_2$N
(P6$_3$/mmc) films have rather strong electron-phonon interaction ($\lambda$ = 0.54)
with a $T_c$ of 4.74 K.

As mentioned before, four crystalline structures (see Fig. 1) have
been reported for the $\beta$-phase Nb$_2$N~\cite{Norlund,Savage,Tera,Zha}.
This is quite unlike other niobium nitrides with different Nb/N ratios.
For example, different structures of NbN were labelled as different phases
(one structure, one phase) (see Table I). Since the physical properties
of a solid depend significantly on the crystalline structure,
the contradicting superconductivity reported for $\beta$-Nb$_2$N
could certainly be attributed to the fact that $\beta$-Nb$_2$N
has several different structures. This has motivated us to
carry out this {\it ab initio} theoretical study on the superconducting properties
of $\beta$-Nb$_2$N in all four reported structures.

By using Allen-Dynes formula (Eq. 3) and the calculated $\lambda$
as well as the other phonon and electron parameters, we estimate the $T_c$ values for all
the considered nitrides, as listed in Table IV.
First of all, we notice that the calculated $T_c$ values for $\gamma$-Nb$_4$N$_3$ (8.48 K) and
$\beta^\prime$-Nb$_4$N$_5$ (15.3 K) agree rather well with the corresponding experimental
values~\cite{Kaiser,Oya,Jena,Merch} (see Table IV).
Second, different structures of $\beta$-Nb$_2$N indeed
have rather different $T_c$ values, ranging from $\sim$0.6 K to 6.1 K (Table IV).
The calculated $T_c$ of $\beta_2$-Nb$_2$N (P$\bar{3}$m1) (6.1 K) is larger than
$\beta_3$-Nb$_2$N (P6$_3$mmc) (3.9 K), $\beta_4$-Nb$_2$N (Pnnm) (2.6 K) and
$\beta_1$-Nb$_2$N (P$\bar{3}$1m) (0.6 K).
These results clearly demonstrate that further experiments measuring the
superconductivity and crystalline structure simultaneously on the same sample
would be needed to clarify the current confusing experimental results for $\beta$-Nb$_2$N.

It is useful to find connections between the superconductivity and
other physical properties of the nitrides. This could be helped
by McMillan-Hopfield formula \cite{McMillan} $\lambda = [\frac{N(E_F)}{<\omega^2>}]\Sigma_i(\frac{<I^2>_i}{M_i})$
where $<I^2>_i$ is the square of the electron-phonon coupling matrix element averaged
over the Fermi surface and $M_i$ is the atomic mass of $i^{th}$ atom.
Also, $<\omega^2> \approx 0.5\Theta_D^2$.
Clearly, this indicates that $\lambda$ and hence $T_c$ would depend on $N(E_F)$
and  would be relatively large for the electronic bands with a high DOS near the Fermi energy.
The calculated DOS spectra shown in Fig. 3, indicate that the Nb $d$-states dominate
the DOS near $E_F$ for all the structures, and therefore would make major contributions to
the electron-phonon coupling and superconductivity. Thus, the calculated $N(E_F)$ per Nb atom
for the considered nitrides are listed in Table IV.
As can be seen from Table IV, for the considered nitrides, roughly, $\lambda$ and $T_c$ are larger
with larger $N(E_F)$ and smaller $\Theta_D$.
For example, $\delta$-NbN has the largest $\lambda$, $T_c$ and $N(E_F)$ but the smallest $\Theta_D$ (Table IV).

\section{CONCLUSION}
By performing systematic \textit{ab initio} calculations based on the DFT and DFPT,
we have investigated the superconductivity, electronic and phononic band structures,
electron-phonon coupling and
elastic constants of all four reported structures of $\beta$-Nb$_2$N as well as
Nb-rich $\gamma$-Nb$_4$N$_3$ and N-rich $\beta^\prime$-Nb$_4$N$_5$. First,
all four structures of $\beta$-Nb$_2$N are found to be superconductors
with $T_c$ ranging from 0.6 K to 6.1 K,
depending on their structure (Table IV). This finding thus clarifies the long standing confusion that
although Nb$_2$N was labelled as the single $\beta$ phase, contradicting $T_c$
values for $\beta$-Nb$_2$N have been reported in previous experiments. Interestingly,
$\gamma$-Nb$_4$N$_3$ and $\beta^\prime$-Nb$_4$N$_5$
are predicted to be superconductors with rather high $T_c$ of 8.5 K and 15.3 K, respectively.
Second, all the calculated elastic constants and phonon frequencies are
positive, thereby showing that all the considered niobium nitride structures
are mechanically and dynamically stable. This suggests that although only $\beta_1$-Nb$_2$N
is found to be the ground state, the other three structures of $\beta$-Nb$_2$N
could be grown in, e.g., the $\beta$-Nb$_2$N films.
Furthermore, the calculated elastic moduli show
that all the niobium nitrides are hard materials with bulk moduli and hardness being comparable to
or even larger than the well-known hard sapphire.
Third, the calculated electronic band structures reveal
that $\beta_3$-Nb$_2$N, $\gamma$-Nb$_4$N$_3$ and $\beta^\prime$-Nb$_4$N$_5$
are topological metals. Specifically, $\beta_3$-Nb$_2$N and $\gamma$-Nb$_4$N$_3$
possess type-I Dirac nodal points whereas $\beta^\prime$-Nb$_4$N$_5$ has type-II Dirac points.
Finally, the calculated electron-phonon coupling strength, superconductivity and
mechanical property of the niobium nitrides are discussed in terms of their
underlying electronic structures and also Debye temperatures.
For example, that $\beta^\prime$-Nb$_4$N$_5$ has the largest $\lambda$ and  highest $T_c$ among the considered niobium
nitrides, could be attributed to its largest DOS at $E_F$.
All these interesting findings indicate that $\beta$-Nb$_2$N,
$\gamma$-Nb$_4$N$_3$ and $\beta^\prime$-Nb$_4$N$_5$ are hard superconductors with nontrivial band
topology and are promising materials for studying fascinating
phenomena arising from the interplay of hardness, superconductivity and nontrivial band topology.

\section*{Acknowledgments}
The authors acknowledge the support from the National Science and Technology Council
and National Center for Theoretical Sciences (NCTS) in Taiwan. The authors are also
grateful to the National Center for High-performance Computing (NCHC) in Taiwan for the computing time.


\end{document}